\documentclass[12 pt]{article}

\usepackage[T1]{fontenc}
\usepackage[latin1]{inputenc}
\usepackage{amssymb}
\usepackage{amsmath}
\usepackage[dvips]{graphicx}
\usepackage{wasysym}
\usepackage{stmaryrd}
\usepackage{verbatim}
\usepackage{natbib}
\usepackage{pstricks}
\usepackage{pst-text}
\usepackage{hhline}
\usepackage{endfloat}
\usepackage{setspace}

\begin{document}
\sloppy

\newtheorem{exemple}{Exemple}
\newtheorem{theoreme}{Théorème}
\newtheorem{proposition}{Proposition}
\newtheorem{lemme}{Lemme}
\newtheorem{definition}{Définition}

\newcommand{\argmax}[1]{\textrm{argmax}_{#1}}
\newcommand{\argmin}[1]{\textrm{argmin}_{#1}}

%
%
%

\noindent
VARIOUS APPROACHES FOR PREDICTING LAND COVER IN MOUNTAIN AREAS

\vskip 5mm

\noindent Nathalie Villa$^{1,\textrm{a}}$, Martin Paegelow$^2$, Maria T. Camacho Olmedo$^3$, Laurence Cornez$^4$, Fr\'ed\'eric Ferraty$^1$, Louis Ferr\'e$^1$ and Pascal Sarda$^1$.
\vskip 2mm
\noindent $^1$ GRIMM, Equipe d'accueil 3686, Université Toulouse Le Mirail, France\\
$^2$ GEODE UMR 5602 CNRS, Université Toulouse Le Mirail, France\\
$^3$ Instituto de desarrollo regional, Universidad de Granada, Spain\\
$^4$ ONERA, Toulouse, France

\noindent $^\textrm{a}$ Corresponding author e-mail: villa@univ-tlse2.fr
\vskip 4mm
\noindent Key Words: polychotomous regression modelling; multilayer perceptron; classification; prediction; comparison.
\vskip 4mm

\noindent ABSTRACT

\par
Using former maps, geographers intend to study the evolution of the land cover in order to have a prospective approach on the future landscape; predictions of the future land cover, by the use of older maps and environmental variables, are usually done through the GIS (Geographic Information System). We propose here to confront this classical geographical approach with statistical approaches: a linear parametric model (polychotomous regression modelling) and a nonparametric one (multilayer perceptron). These methodologies have been tested on two real areas on which the land cover is known at various dates; this allows us to emphasize the benefit of these two statistical approaches compared to GIS and to discuss the way GIS could be improved by the use of statistical models.
\vskip 5mm

\noindent 1.   PREDICTING LAND COVER

\par
From the sketch maps made by geographers or from the analysis of satellite images or aerial photographs, we can build land cover maps for a given country which can be rather precise: the studied area is then cut into several squared pixels whose sides are about 20 meters long and whose land cover is known on various dates. The type of land cover can be chosen from a pre-determined list: coniferous forests, deciduous forests, scrubs, \ldots

Here, we are not interested in making such maps (for satellite data analysis, see (Cardot {\it et al.}, 2003)). Our purpose is to contruct a simulated land cover map at a given future date, by the use of land cover maps at older dates and of other environmental variables; on a geographical point of view, prospective simulations have a great interest to help the local administrations to develop these mountain areas. The idea is then to compare different approaches in order to confront their ability to be generalized to various mountain areas.

For a given pixel, determined by its spatial coordinates, latitude ($i$) and longitude ($j$), the value of the land cover on date $t$, $c_{i,j}(t)$, is a categorical random variable depending on several variables:
\begin{itemize}
\item[$\bullet$]  the land cover of this pixel on previous dates: $c_{i,j}(t-1),\ldots,c_{i,j}(t-T)$ (\emph{time serie of length $T$});
\item[$\bullet$]  the land covers of the neighbouring pixels on previous dates: $V_{i,j}(t-1),\ldots, V_{i,j}(t-T)$, where $V_{i,j}(t-\tau)$ is a set of values of land cover on date $t-\tau$ for the pixels in a neighbourhood of the pixel $(i,j)$ (\emph{vectorial time serie});
\item[$\bullet$]  some environmental variables: for example, the elevation, the aspect, the proximity of roads and villages, \ldots: $Y^1_{i,j},\ldots,Y^p_{i,j}$.
\end{itemize}

We face here a problem of classification in which the predictors are both qualitative and quantitative and are also highly dependent (spatial time process). To solve this question, we propose to use and to compare two well-known statistical approaches with the empirical geographic method (namely the GIS, Geographic Information System). The first of these methods is a generalized linear model in which we estimate the parameters of the model by maximizing a log-likelihood type criterion. The second one uses a supervised multilayer perceptron. By confronting these various approaches, we expect to give ideas in order to improve the GIS approach.

A comparison of these two approaches was done on two little areas: the ``Garrotxes'' (``Pyrénées Orientales'', south west of France) and the ``Alta Alpujarra Granaderia'' (Sierra Nevada, Spain) where several surveys of the land cover were done at various dates. We confronted the various scenarii constructed with the real maps.

In the following, we describe the data more precisely (section 2) and present the two approaches (section 3). Then we present how we applied these methodologies on these data sets (section 4) and finally, we compare the results obtained by analyzing the advantages and the limits of the models (section 5).
\vskip 5mm

\noindent 2.   DESCRIPTION OF THE DATA SETS

\par
The areas under study stand in the moutains ``Pyrénées'' for the Garrotxes and Sierra Nevada ``Alta Alpujarra''. A big drift from the land has led to the desertion of the land under cultivation and the recovery of the fields by scrubs and forests. There is almost none human action on these areas. The aridity of the climate explains a much slower dynamic in the spanish area than in the Garrotxes: we count 3 times less pixels changing in the Alta Alpujarra than in the Garrotxes. On the contrary, the french area is considered, at least on a geographical point of view, as a dynamic area and it is then more difficult to predict the land cover.

We are given quantitative and qualitative informations through maps divided into pixels: about 241~000 pixels for the French area and 560~000 for the Spanish one (which is much bigger). For each pixel, we know:

\begin{itemize}
\item[$\bullet$] a categorical variable which is the land cover at different dates: 3 dates (1980, 1990 and 2000) were avalaible for the Garrotxes and 4 dates (1957, 1974, 1987 and 2001) for the Alta Alpujarra. As the land cover evolution is very slow in the Sierra Nevada (less than 25\% of the pixels had changed their value between 1957 and 2001), these dates were considered as equidistant, according to geographers opinion. This categorical variable was taken from a list of several choices (8 for the Garrotxes and 9 for the Alta Alpujarra) which are of classical use in geography. These data were used to make maps of the studied area (see Figure \ref{Figure1});
\begin{figure}[h]
\begin{center}
\begin{tabular}{c}
\makebox[5 cm][l]{\includegraphics[width=4.5 cm]{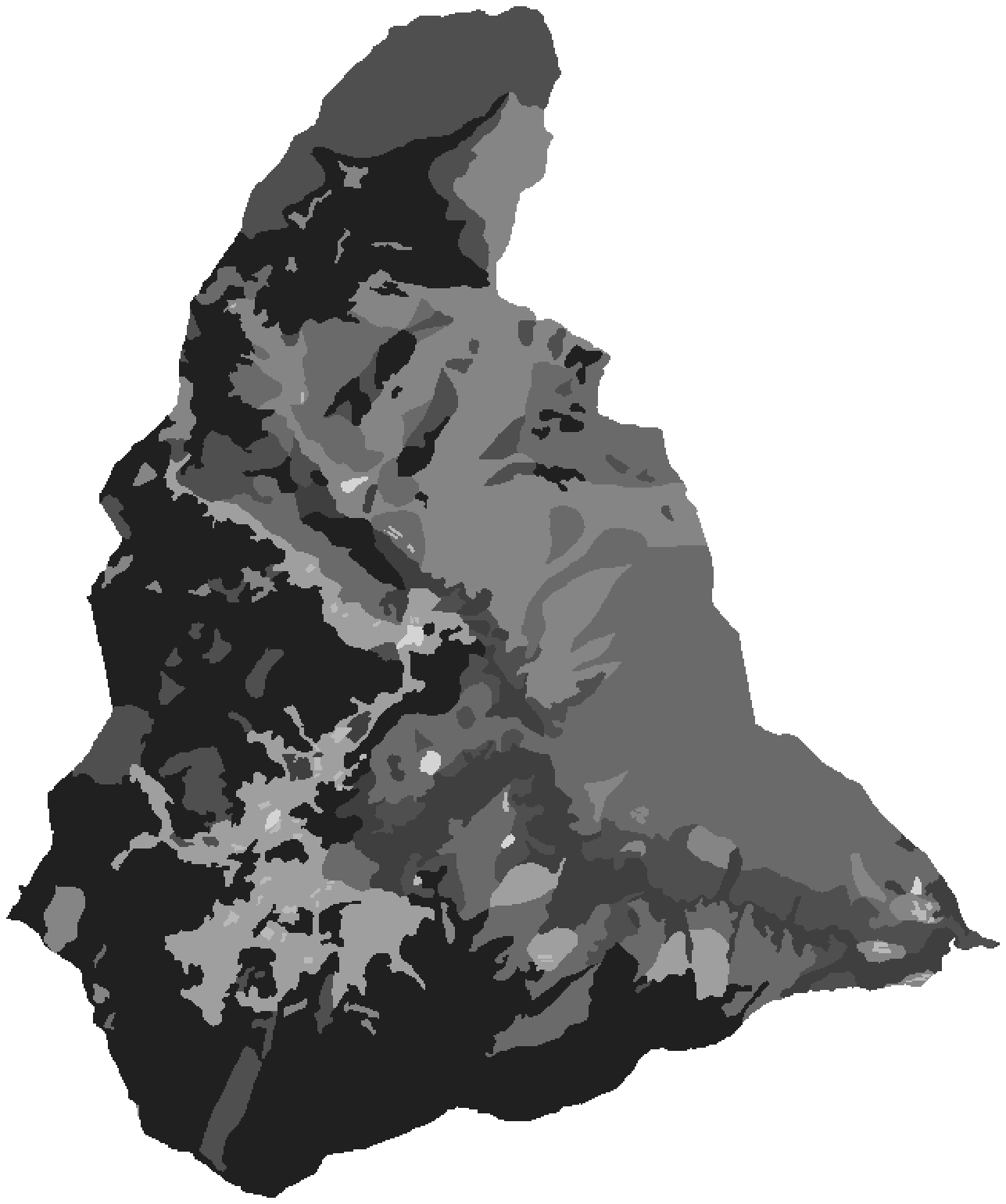}}
\makebox[2 cm]{ }
\makebox[5 cm][r]{\includegraphics[width=5 cm]{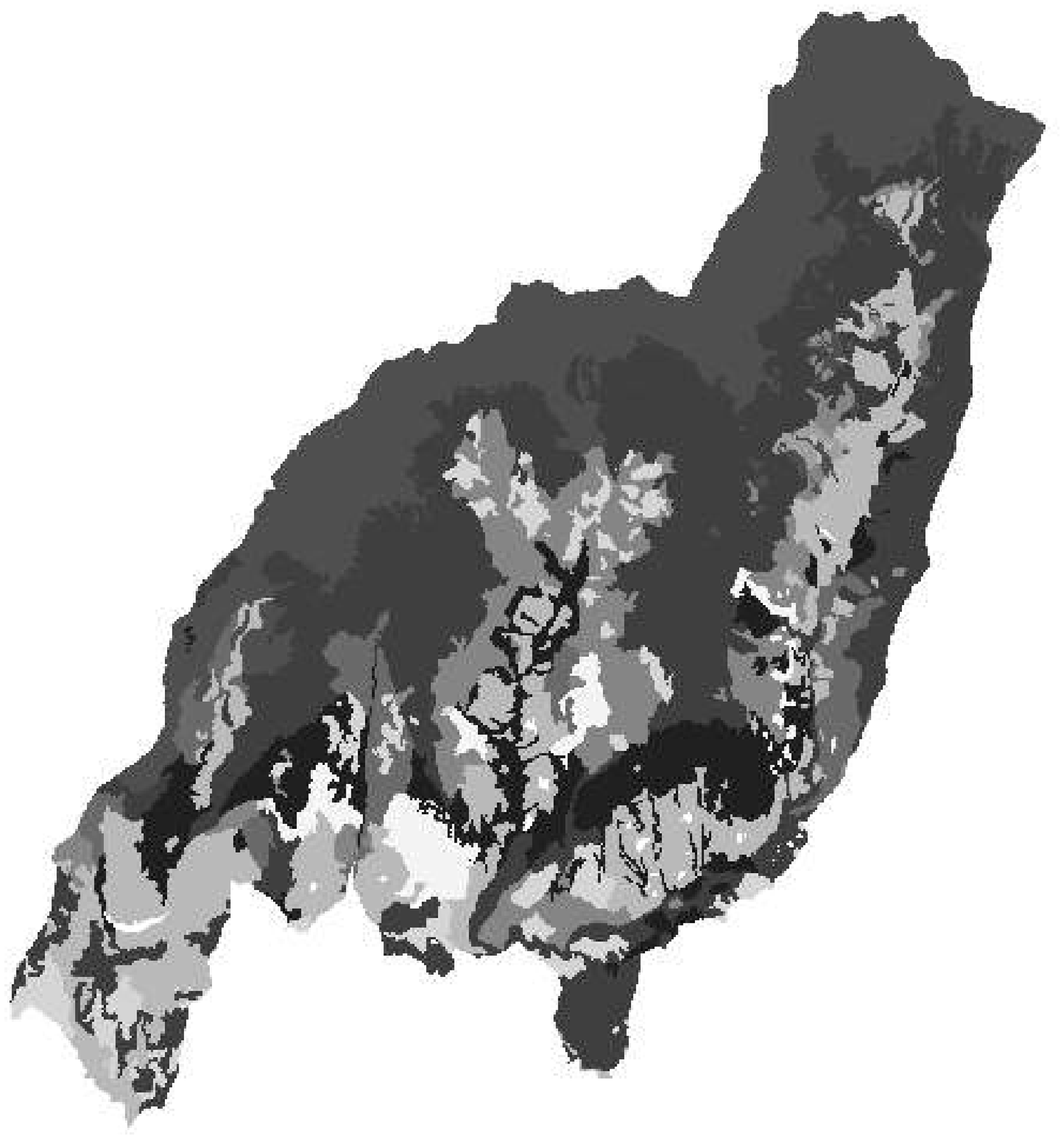}}\\
\makebox[4.5 cm][c]{\includegraphics[width=3 cm]{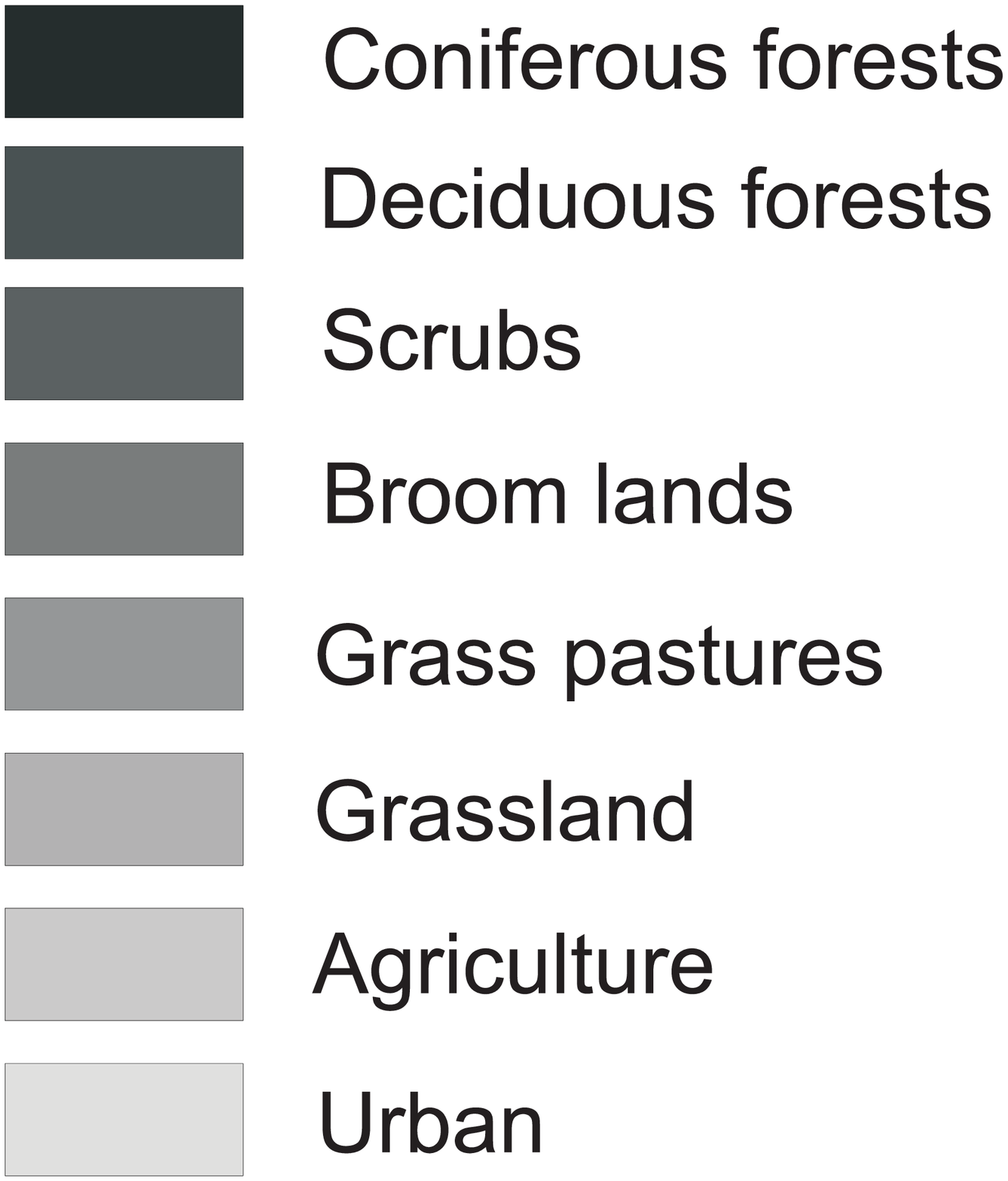}}
\makebox[2 cm]{ }
\makebox[4.5 cm][c]{\includegraphics[width=4.5 cm]{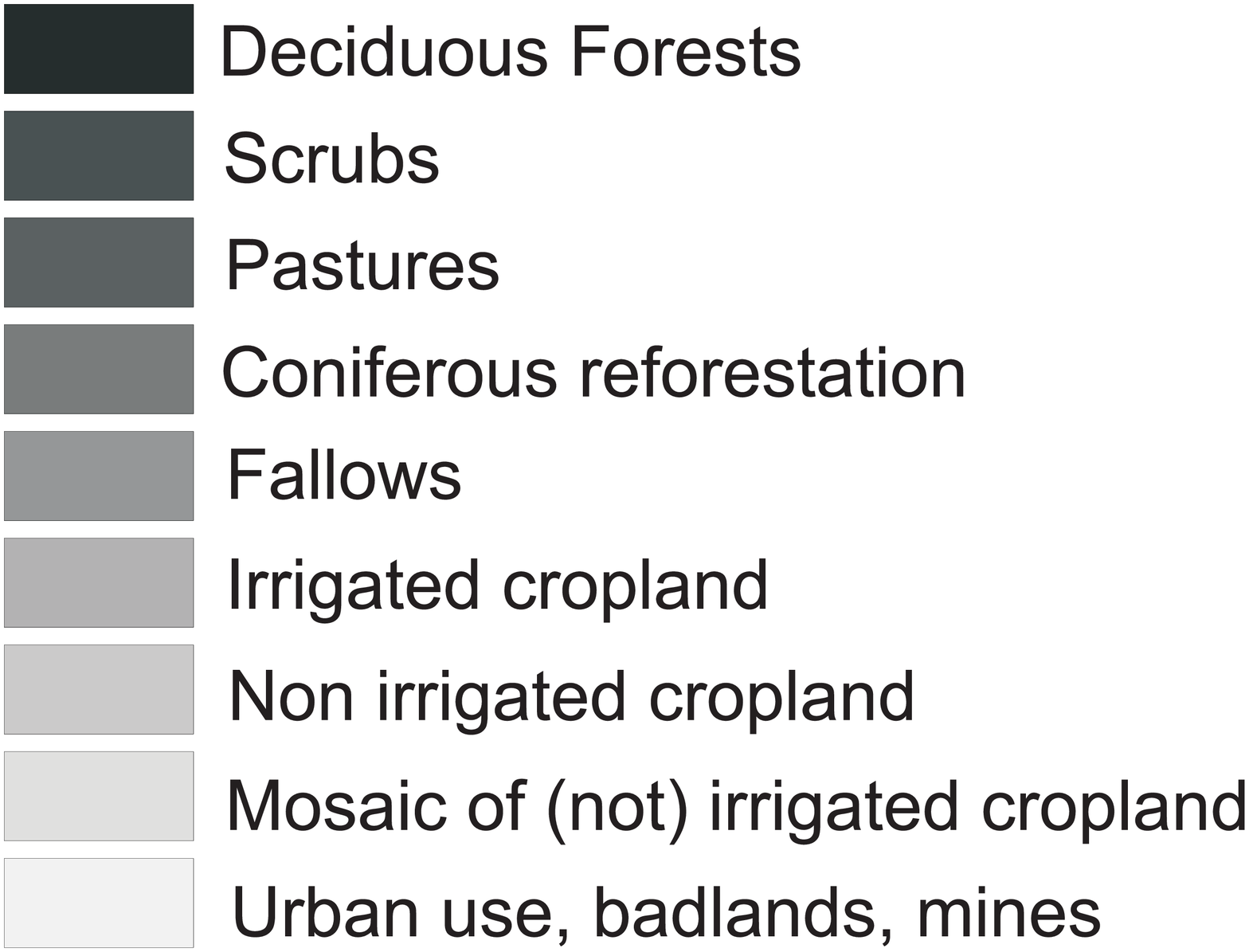}}
\end{tabular}
\end{center}
\caption{Land cover for the Garrotxes (1980 - left) and for the Alta Alpujarra (1957 - right)}
\label{Figure1}
\end{figure}

\item[$\bullet$]  several environmental variables; some of them are of numeric type (the elevation, the slope, the aspect, the distance of roads and villages,\ldots) and others are of categorical type (forest and pasture management: governmental or not ? ground geological type, \ldots). The environmental variables were not the same for the Garrotxes and the Alta Alpujarra (see Figure \ref{Figure2} for examples of environmental variables); all these environmental variables kept the same value at all dates.
\begin{figure}[h]
\begin{center}
\makebox[5 cm][l]{\includegraphics[width=5 cm]{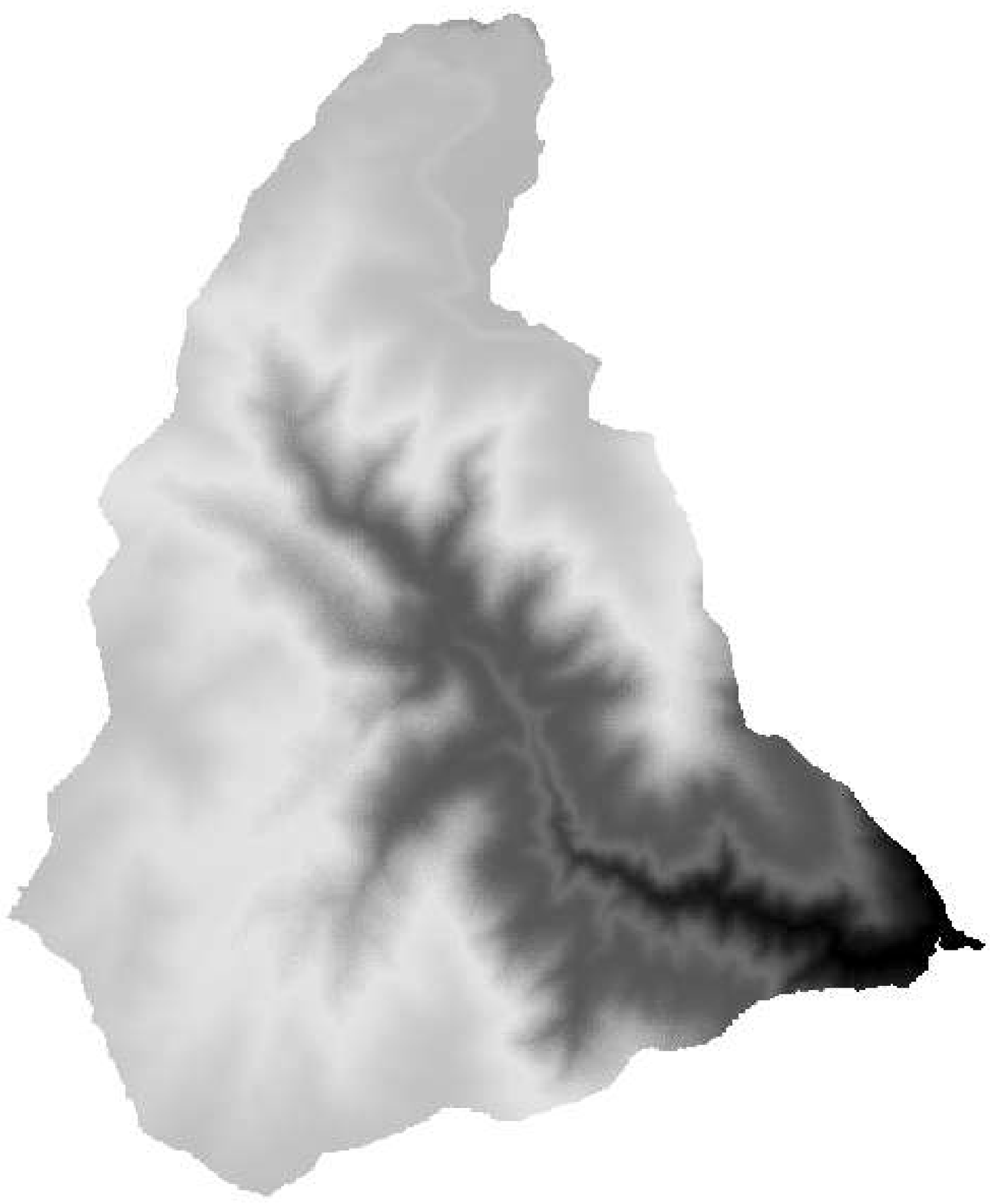}}
\makebox[1 cm]{ }
\makebox[5 cm][l]{\includegraphics[width=5 cm]{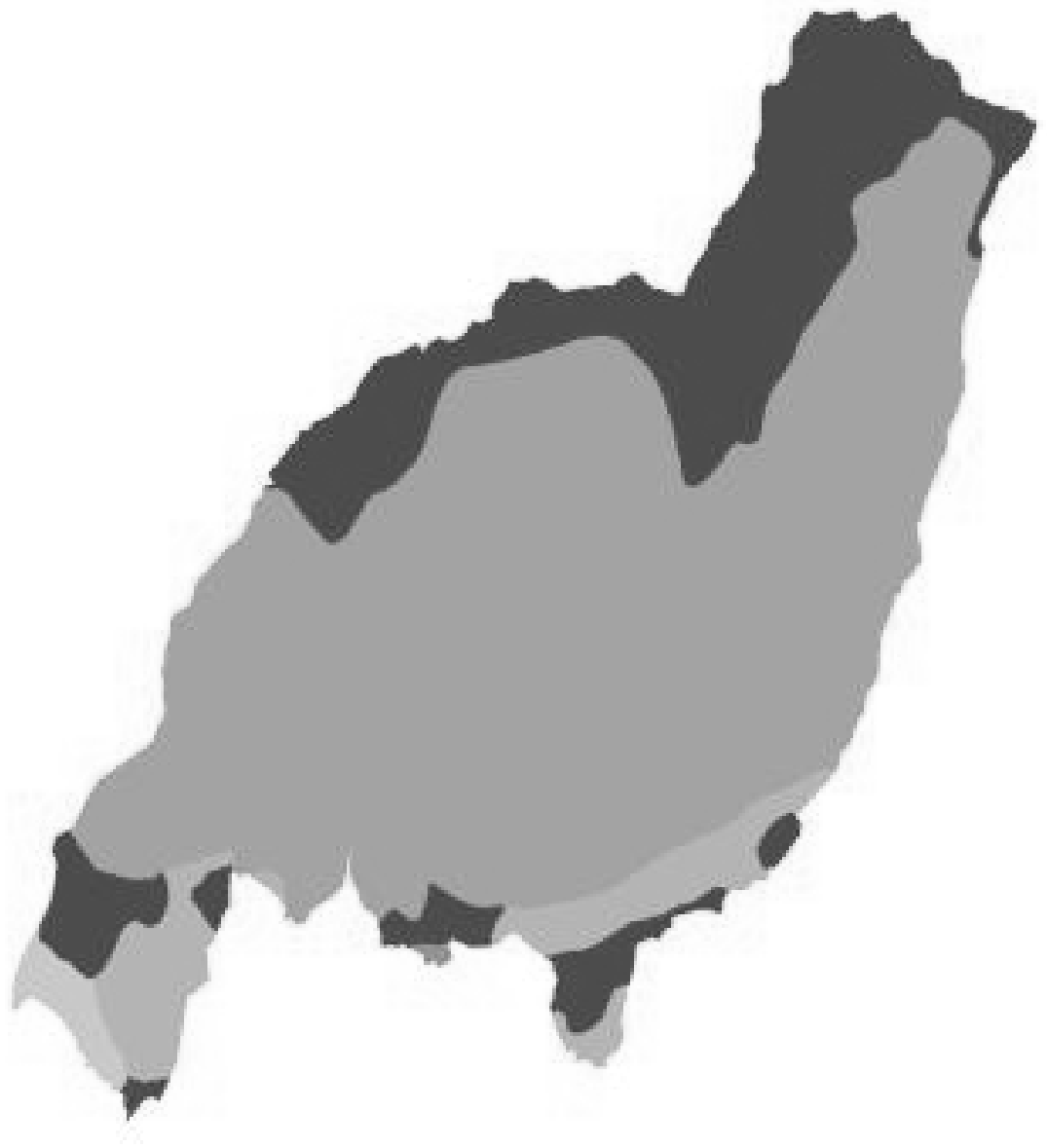}}
\caption{Examples of a numerical variable (elevation for the Garrotxes - left) and a categorical one (ground geological type for the Alta Alpujarra - right)}
\label{Figure2}
\end{center}
\end{figure}
\end{itemize}
\vskip 5mm

\noindent 3.   PRESENTATION OF THE TWO APPROACHES

\par
Geographers usually estimate the land cover evolution by an empirical method which allows to introduce some expert knowledge. The so-called GIS (Geographic Information System) approach is time expensive and necessitates precise knowledge on the geographic constraints of the area under study. Roughly speaking, the method consists in two steps: at first one computes time transition probabilities for each land cover type whereas, in a second step, one uses spatial constraints (introduced by an expert) for ``smoothing'' the maps obtained at the first step (see (Paegelow {\it et al.}, 2004) or (Paegelow and Camacho Olmedo, 2005) for further details on GIS for these data sets). In order to propose automatic alternatives to the GIS, which can take in the same model the spatio-temporal nature of the problem, two approaches have been adapted to estimate the evolution of the land cover: the first one, polychotomous regression modelling, is a generalized linear approach based on the maximum log-likelihood method. The second one, multilayer perceptron, is a popular method which has recently proved its great efficiency to solve various types of problems. 

The idea is to confront a parametric linear model with a non parametric one to provide a collection of automatic statistical methods for geographers. They both have concurrent advantages that have to be taken into account when choosing one of them: the polychotomous regression modelling is faster to train than multilayer perceptrons, especially in high dimensional spaces and does not suffer from the existence of local minima. On the contrary, multilayer perceptrons can provide nonlinear solutions and are then more flexible than the linear modelling; moreover, both methods are easy to implement even for non statisticians through the pre-made softwares (for example, ``Neural Network'' Toolbox for neural network with Matlab).
\vskip 5mm

\noindent 3.1.   THE MODEL

\par
Let us now describe the statistical setting more formally. We note $X_{i,j}(t)$ the vector of variables that could explain the value of the land cover for a given pixel $(i,j)$ on date $t$. We suppose that the time dependence is of order 1; then, $X_{i,j}(t)$ contains:
\begin{itemize}
\item[$\bullet$]  \emph{for the time series:} the value of the land cover for the pixel $(i,j)$ at the previous time $t-1$;
\item[$\bullet$]  \emph{for the spatial aspect:} the frequency of each type of land cover in the neighbourhood of pixel $(i,j)$ on the previous date. Then, the shape and the size of the neighbourhood had to be chosen. For the shape, we had many choices: the simpler one was a square neighbourhood or a star-shaped neighbourhood around the pixel $(i,j)$; the most sophisticated could use the slope to better take into account the morphological influences of the land. For the size of the neighbourhood, we had to find at which distance a pixel could influence the land use of pixel $(i,j)$. Moreover, for the multilayer perceptrons, in order to respect the spatial aspect of the problem, we weighted the influence of a pixel by a decreasing function of its distance to the pixel $(i,j)$ (see Figure \ref{Figure3}).

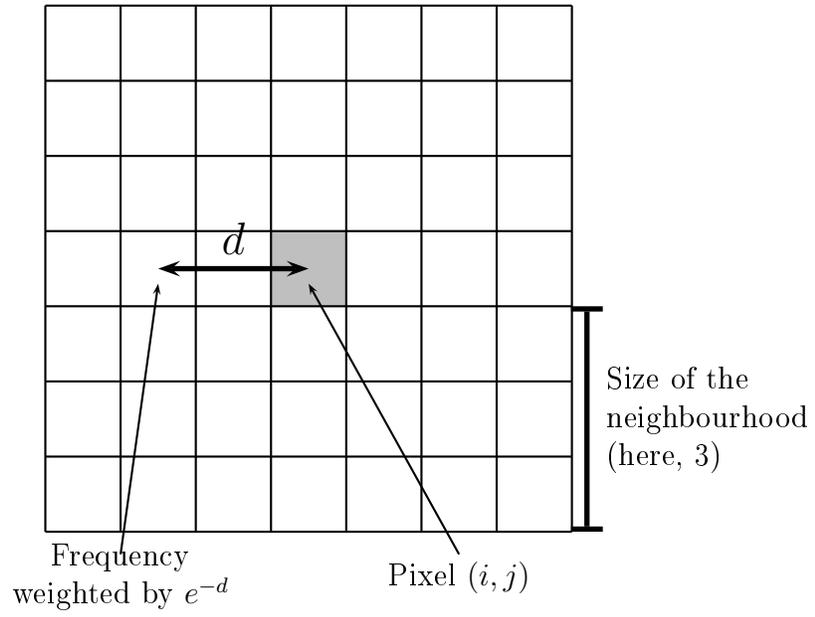
\begin{figure}[h]
\begin{center}
\begin{pspicture}(0,0)(12,9)
\pspolygon*[linecolor=lightgray](4,4)(5,4)(5,5)(4,5)
\psline(1,1)(1,8)
\psline(2,1)(2,8)
\psline(3,1)(3,8)
\psline(4,1)(4,8)
\psline(5,1)(5,8)
\psline(6,1)(6,8)
\psline(7,1)(7,8)
\psline(8,1)(8,8)
\psline(1,1)(8,1)
\psline(1,2)(8,2)
\psline(1,3)(8,3)
\psline(1,4)(8,4)
\psline(1,5)(8,5)
\psline(1,6)(8,6)
\psline(1,7)(8,7)
\psline(1,8)(8,8)
\psline[arrowsize=6pt,linewidth=2pt]{<->}(4.5,4.5)(2.5,4.5)
\psline[arrowsize=6pt,linewidth=2pt]{|-|}(8.2,1)(8.2,4)
\psline[arrowsize=3pt]{->}(2,0.7)(2.5,4.3)
\psline[arrowsize=3pt]{->}(6.5,0.7)(4.5,4.3)
\rput(2,0.4){\begin{tabular}{c} Frequency\\weighted by $e^{-d}$\end{tabular}}
\rput(6.5,0.4){Pixel $(i,j)$}
\rput(9.8,2.5){\begin{tabular}{l} Size of the\\ neighbourhood\\ (here, 3) \end{tabular}}
\rput(3.5,4.9){{\Large $d$}}
\end{pspicture}
\caption{An example of neighbourhood}
\label{Figure3}
\end{center}
\end{figure}

\item[$\bullet$] \emph{environmental variables} (slope, elevation, \ldots).
\end{itemize}

Let us repeat that $c_{i,j}(t)$ is the land cover for a given pixel on date $t$. We note ${\cal C}_1,\ldots,{\cal C}_K$ the different types of land cover. Then, for every $k=1,\ldots,K$, we try to estimate the probability $P(c_{i,j}(t)={\cal C}_k|X_{i,j}(t))$ that the pixel $(i,j)$ has a land cover equal to ${\cal C}_k$ given the vector $X_{i,j}(t)$; thus, the model is of the following form :
\begin{equation}
\label{geo1_modele_nnet}
P(c_{i,j}(t)={\cal C}_k|X_{i,j}(t))=f_k(X_{i,j}(t)).
\end{equation}
Once a model was chosen through $f_k$, these probabilities were estimated by the way of a multi-layer perceptron or a generalized linear model and we predicted the type of land cover, $c_{i,j}(t)$, by the rule of maximum:
\[\argmax{k=1,\ldots,K} P(c_{i,j}(t)={\cal C}_k|X_{i,j}(t)).\]
In both approaches, we estimated $f_k$ thanks to a training sample. To that end, we have collected the values of the predictors and of the land cover for many pixels on various dates (see next section for more details); the observations are denoted by $(X^{(1)},c^{(1)}),\ldots,(X^{(N)},c^{(N)})$.

The time and spatial aspects are taking into account together both by the polychotomous regression modelling and by the multilayer perceptron and the land cover prediction is performed in a single estimation procedure. This is not the case for the usual GIS approach which is performed in two steps: it first estimates the land cover probability by modelling a time serie and it then introduces a spacial smoothing with environmental constraints.
\vskip 5mm

\noindent 3.2.   POLYCHOTOMOUS REGRESSION MODELLING

\par
When we wish to predict a categorical response given a random vector, a useful model is the {\em multiple logistic regression} (or {\em polychotomous regression}) model (Hosmer and Lemeshow, 1989). A smooth version of this kind of method can be found in (Kooperberg {\it et al.}, 1997). Applications of these statistical techniques to several situations such as in medicine or for phoneme recognition can be found in these two works. Their good behaviour both on theoretical and practical grounds have been emphasized. In our case, where the predictors are both categorical and scalar, we then have the derived model below.

Let us note, for $k=1,\ldots,K$ 
\[
\theta\left({\cal C}_k|X_{i,j}(t)\right)=\log\frac{P\left(c_{i,j}(t)={\cal C}_k|X_{i,j}(t)\right)}{P\left(c_{i,j}(t)={\cal C}_K|X_{i,j}(t)\right)}.
\]
Then, we get the following expression
\begin{equation}
\label{geo1_eq_poly}
P\left(c_{i,j}(t)={\cal C}_k|X_{i,j}(t)\right)=\frac{\exp \theta\left({\cal C}_k|X_{i,j}(t)\right)}{\sum_{k'=1}^K\exp \theta\left({\cal C}_{k'}|X_{i,j}(t)\right)}.
\end{equation}
Now, to estimate these conditional probabilities, we use the parametric approach to the polychotomous regression problem, that is the linear model
\begin{equation} \label{geo1_model_fp}
\theta\left({\cal C}_k|X_{i,j}(t)\right)=\alpha_k+\sum_{c \in V_{i,j}(t-1)}\sum_{l=1}^K \beta_{kl}1\!\!1_{[c={\cal C}_l]}+\sum_{r=1}^p\gamma_{kr}Y_{i,j}^r,
\end{equation}
where we recall that $V_{i,j}(t-1)$ are the values of the land cover in the neighbourhood of the pixel $(i,j)$ on the previous date $t-1$ and $(Y_{i,j}^r)_r$ are the values of the environment variables. Let us call ${\delta}=(\alpha_1,\ldots,\alpha_{K-1},\beta_{1,1},\ldots,\beta_{1,K},\beta_{2,1},$\\$\ldots,\beta_{2,K},\ldots,\beta_{K-1,1},\ldots,\beta_{K-1,K},\gamma_{1,1},\ldots,\gamma_{1,K},\ldots,\gamma_{K-1,1},\ldots,\gamma_{K-1,p})$, the parameters of the model to be estimated. We have to notice that since $\theta\left({\cal C}_K|X_{i,j}(t)\right)=0$, we have $\alpha_K=0$, $\beta_{K,l}=0$ for all $l=1,\ldots,K$, and $\gamma_{K,r}=0$ for all $r=1,\ldots,p$. We now have to estimate the vector of parameters ${\delta}$. For that end, we use a penalized likelihood estimator which is performed on the training sample. Let us write the penalized log-likelihood function for model (\ref{geo1_model_fp}). It is given by
\begin{equation}\label{geo1_penloglik}
l_{\varepsilon}(\delta)\ =\ l(\delta)-\varepsilon\sum_{n=1}^N\sum_{k=1}^K u_{nk}^2,
\end{equation}
where the log-likelihood function is
\begin{equation}
\label{geo1_eq_poly2}
l(\delta)=\log \left(\prod_{n=1}^N P_\delta\left(c^{(n)}|X^{(n)}\right)\right).
\end{equation}
In this expression, $P_\delta(c^{(n)}|X^{(n)})$ is the value of the probability given by (\ref{geo1_eq_poly}) and (\ref{geo1_model_fp}) for the observations $(X^{(n)},c^{(n)})$ and the value $\delta$ of the parameter.

In expression (\ref{geo1_eq_poly2}), $\varepsilon$ is a penalization parameter and, for $k=1,\ldots,K$,\\  $\displaystyle u_{nk}=\theta_\delta({\cal C}_k|X^{(n)})-\frac{1}{K}\sum_{k'=1}^K\theta_\delta({\cal C}_{k'}|X^{(n)})$. Our penalized likelihood estimator $\widehat{{\delta}}_{\varepsilon}$ satisfies:
\[
\widehat{{\delta}}_{\varepsilon} = \argmax{\delta\in\ \mathbb{R}^M} l_{\varepsilon}(\delta),
\]
where $M=K^2+(K-1)*p-1$ denotes the number of parameters to be estimated.

As pointed out by (Kooperberg {\it et al.}, 1997) in the context of smooth polychotomous regression, it is possible that, without the penalty term, the maximization of the log-likelihood function $l(\delta)$ leads to infinite coefficients $\beta_{k,l}$. In our model it may be the case, for example, when, for fixed $k$, the value of the predictor is equal to zero for all $(i,j)$. Actually, this ``pathological'' case cannot really occurs in practice but for classes $k$ with a few number of members, the value of the predictor is low and then a numerical unstability happens when maximizing the log-likelihood. Then, the form of the penalty based on the difference between the value $\theta_\delta(\mathcal{C}_k|X^ {(n)})$ for class $k$ and the mean over all the classes has the aim of preventing this unstability by forcing $\theta_\delta(\mathcal{C}_k|X^ {(n)})$ to be not too far from the mean. On another side, for reasonable values of $\epsilon$, we can expect that the penalty term does not affect so much the estimation of parameters while it guarantees numerical stability. Finally, numerical maximization of the penalized log-likelihood function is achieved by a Newton-Raphson algorithm.
\vskip 5mm

\noindent 3.3.   MULTILAYER PERCEPTRON

\par
Neural networks have a great adaptability to any statistical problems and especially to overcome the difficulties of non linear problems even if the predictors are highly correlated; thus it is not surprising to find them used in the chronological series prediction ((Bishop, 1995),  (Lai and Wong, 2001) and (Parlitz and Merkwirth, 2000)). The main interest of neural networks is their ability to approximate any function with the desired precision (universal approximation): see, for instance, (Hornik, 1991).

\begin{figure}[h]
\begin{center}
\begin{pspicture}(0,0)(12,5)
\pscircle(1,2.5){1}
\pscircle(9,2.5){1}
\psellipse(5,4)(1.5,1)
\psellipse(5,1)(1.5,1)
\pscircle*(5,2.2){0.05}
\pscircle*(5,2.5){0.05}
\pscircle*(5,2.8){0.05}
\psline[arrowsize=6pt]{->}(2,2.5)(3.5,4)
\psline[arrowsize=6pt]{->}(2,2.5)(3.5,1)
\psline[arrowsize=6pt]{->}(6.5,4)(8,2.5)
\psline[arrowsize=6pt]{->}(6.5,1)(8,2.5)
\psline[arrowsize=3pt]{->}(4,2.8)(4.9,2.8)(4.9,3)
\psline[arrowsize=3pt]{->}(3.5,0)(4.3,0)(4.3,0.2)
\psline[arrowsize=12pt,linewidth=2pt]{->}(10,2.5)(10.7,2.5)
\pscurve(5,3.4)(5.2,3.45)(5.4,4.15)(5.6,4.2)
\pscurve(5,0.4)(5.2,0.45)(5.4,1.15)(5.6,1.2)
\rput(1,2.5){{\Large $X$}}
\rput(9,2.5){{\Large $\sum$}}
\rput(2.55,3.7){{\large $w^{(1)}$}}
\rput(7.5,3.7){{\large $w^{(2)}$}}
\rput(5,4.5){\textsc{Neuron 1}}
\rput(4.5,3.7){$\sum +$}
\rput(4.5,0.7){$\sum +$}
\rput(5,1.5){\textsc{Neuron $q_2$}}
\rput(3.5,2.8){$w^{(1)}_{1,0}$}
\rput(3,0){$w^{(1)}_{q_2,0}$}
\psset{linestyle=none}
\pstextpath[c]{\psline(11,0)(11,5)}{OUTPUT}
\end{pspicture}
\caption{Multilayer perceptron with one hidden layer}
\label{Figure4}
\end{center}
\end{figure}
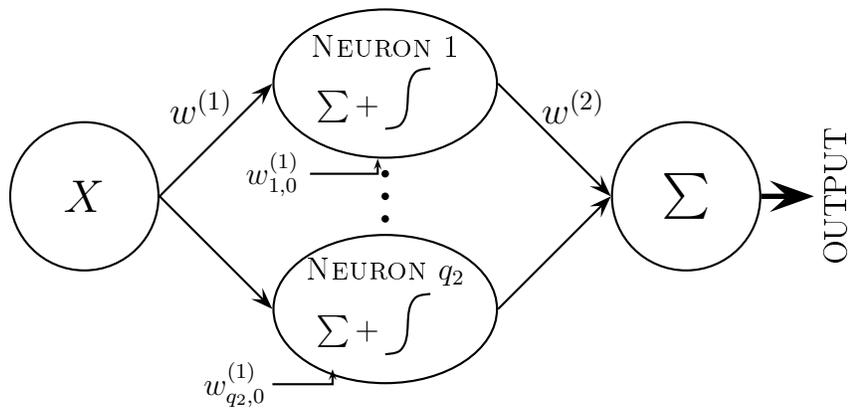

Here we propose to estimate, in model (\ref{geo1_modele_nnet}), the function $f_k$ in the form of a multilayer perceptron with one hidden layer (see Figure \ref{Figure4}), $\psi$, which is a function from $\mathbb{R}^q$ to $\mathbb{R}$ that can be written, for all $x$ in $\mathbb{R}^q$, as
\[\psi_w(x)=\sum_{i=1}^{q_2} w_i^{(2)} g\left(\langle x,w_i^{(1)} \rangle+w_{i,0}^{(1)}\right),\]
where $q_2$ in $\mathbb{N}$ is the number of neurons on the hidden layer, $(w_i^{(1)})_{i=1,\ldots,q_2}$ (respectively $(w_i^{(2)})_{i=1,\ldots,q_2}$, $(w_{i,0}^{(1)})_{i=1,\ldots,q_2}$) are in $\mathbb{R}^q$ (resp. $\mathbb{R}$) and are called weights of the first layer (resp. weights of the second layer, bias) and where $g$, the activation function, is a sigmoïd; for example, $g(x)=\frac{1}{1+e^{-x}}$.

Then, the output of the multilayer perceptron is a smooth function (here it is indefinitly continuous and derivable) of its input. This property ensures that the neural network took into account the spatial aspect of the data set, since two neighbouring pixels have ``close'' values for their predictor variables.

To determine the optimal value for weights $w=((w_i^{(1)})_i,(w_i^{(2)})_i,(w_{i,0}^{(1)})_i)$, we minimized, as it is usual, the quadratic error on the training sample: for all $k=1,\ldots,K$, we chose
\begin{equation}
\label{geo1_poids_nn}
w_{opt}^k=\argmin{w\in \mathbb{R}^{q_2(q+2)}} \sum_{n=1}^N \left[c^{(n)}_k-\psi_w^k(X^{(n)})\right]^2,
\end{equation}
where $c^{(n)}$ and the categorical data in $X^{(n)}$ are written on a disjunctive form. This can be performed by classical numerical methods of the first or the second order (such as gradient descent or conjugate gradients, \ldots) but faces local minima problems. We explain in section 4 how we overcome this difficulty. Finally, (White, 1989) gives many results that ensure the convergence of the optimal empirical parameters to the optimal theoretical parameters.
\vskip 5mm

\noindent 4.   PRACTICAL APPLICATION TO THE DATA SETS

\par
In order to compare the two approaches, we applied the same methodology: we first determined the optimal parameters for each approach (training step, see below) and then, we used the first maps to predict the last one and compared the errors to real map (comparison step, see section~5).

As usual in statistical methods, there are two stages in the training step: the \emph{estimation step} and the \emph{validation step}.
\begin{itemize}
\item[$\bullet$] The \emph{estimation step} consists in estimating the parameters of the models (either for the polychotomous regression or the neural network);
\item[$\bullet$] The \emph{validation step} allows us to choose, for both methodologies, the best neighbourhood, for polychotomous regression, the penalization parameter and, for neural network, the number of neurons on the hidden layer. Concerning the neighbourhood, we only considered square shapes so choosing a neighbourhood is equivalent, in our procedure, to determine its size.
\end{itemize}

\begin{figure}[h]
\begin{center}
\includegraphics[width=5 cm]{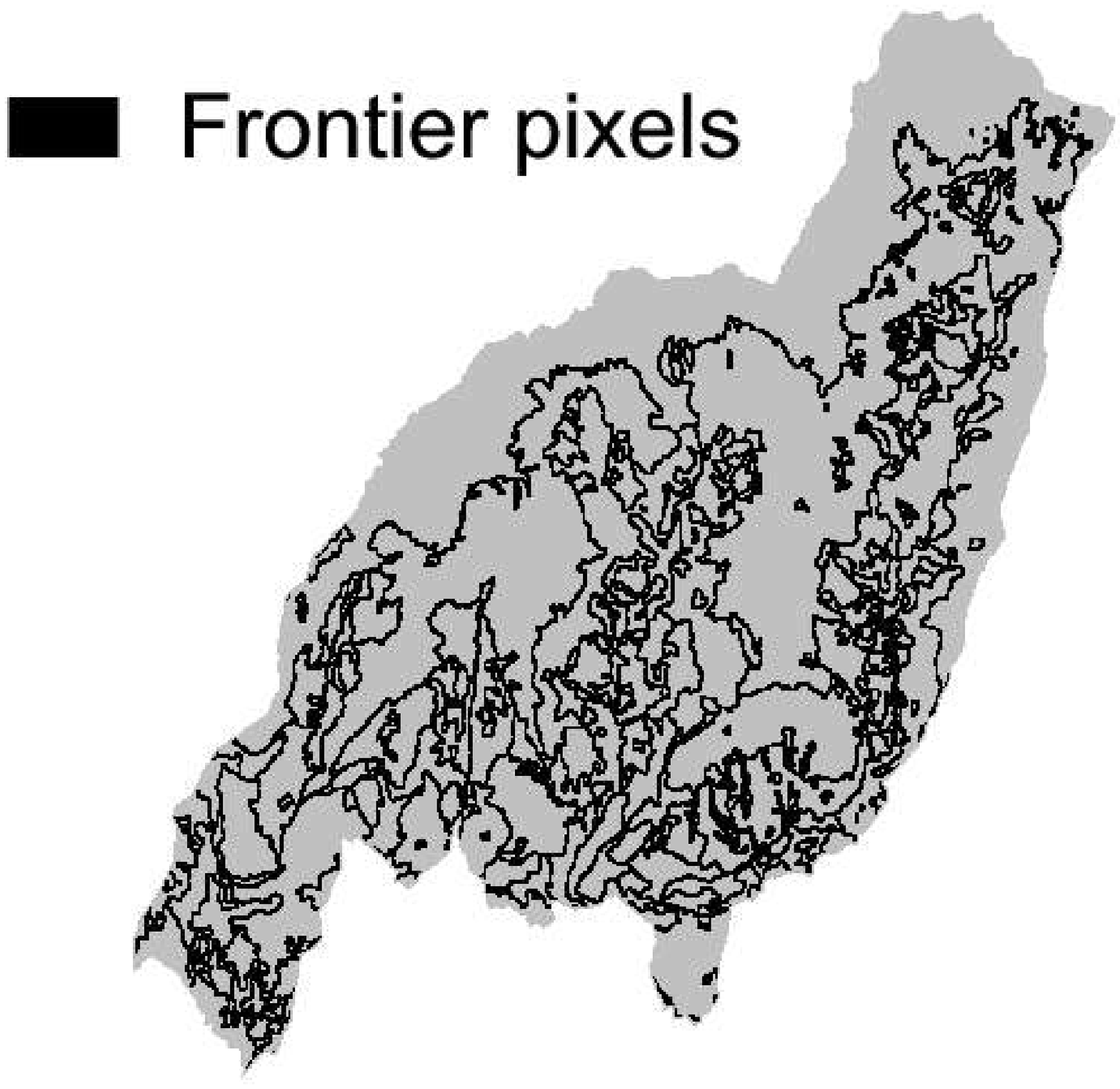}
\end{center}
\caption{Frontier pixels (order 4) for the 1957 map (Alta Alpujarra)\label{Figure5}}
\end{figure}

For the Sierra Nevada, we saw that large areas are constant, thus we only used the pixels for which one neighbour, at least, has a different land cover. These pixels are called ``frontier pixels''; the others were considered as constant (see Figure \ref{Figure5}). For the generalized linear model, we used the whole frontier pixels of the 1957/1974 maps for the estimation set and the whole 1974/1987 maps for the validation set. We then constructed the estimated 2001 map from the 1987 one. For the multilayer perceptron, we reduced the training set size in order not to have huge computational times when minimizing the loss function. Then, estimation and validation data sets were chosen randomly in the frontier pixels of the 1957/1974 and 1974/1987 maps.

For the Garrotxes data set, due to the fact that we only had got 3 maps and much less pixels, we had to use the 1980/1990 maps for the estimation step (only their frontier pixels for the MLP) and the whole 1990/2000 ones for the validation step. This led to a biased estimate when constructing the 2000 map from the 1990 map but, as our purpose is to compare two models and not to make significant the error rate, we do not consider this bias as important.
\vskip 5mm

\noindent 4.1.   POLYCHOTOMOUS REGRESSION

\par
\begin{itemize}
\item[$\bullet$] The {\em estimation step} produces the estimated parameter vector $\widehat{{\delta}}_{\varepsilon}$ of the parameters ${{\delta}}_\varepsilon$ of model (\ref{geo1_model_fp}) for given neighbourhood and penalization parameter $\varepsilon$. This step was repeated for various values concerning both neighbourhood and penalization parameter.
\item[$\bullet$] {\em Validation step:} Once given an estimated parameter vector \linebreak $\widehat{{\delta}}_{\varepsilon}=(\widehat{\alpha}_1,\ldots,\widehat{\alpha}_{K-1},\widehat{\beta}_{1,1},\ldots,\widehat{\beta}_{1,K},\widehat{\beta}_{2,1},\ldots,\widehat{\beta}_{2,K},\ldots,\widehat{\beta}_{K-1,1},\ldots,\widehat{\beta}_{K-1,K},$\\$\widehat{\gamma}_{1,1},\ldots,\widehat{\gamma}_{1,p},\ldots,\widehat{\gamma}_{K-1,1},\ldots,\widehat{\gamma}_{K-1,p})$, the quantities 
\[
\widehat{P}\left(c_{i,j}(t)={\cal C}_k|X_{i,j}(t)\right)=\frac{\exp \widehat{\theta}\left({\cal C}_k|X_{i,j}(t)\right)}{\sum_{k'=1}^K\exp \widehat{\theta}\left({\cal C}_{k'}|X_{i,j}(t)\right)},
\]
were calculated, for all $k=1,\ldots,K$, with
\[
\widehat{\theta}\left({\cal C}_k|X_{i,j}(t)\right)=\widehat{\alpha}_k+\sum_{c\in V_{i,j}(t)}\sum_{l=1}^K\widehat{\beta}_{kl}1\!\!1_{[c={\cal C}_l]}+\sum_{r=1}^p\widehat{\gamma}_{kr}Y_{i,j}^r.
\]
At each pixel $(i,j)$ for the predicted map on date $t$, we affected the most probable vegetation type namely the ${\cal C}_k$ which maximizes 
\[
\left\{\widehat{P}\left(c_{i,j}(t)={\cal C}_k|X_{i,j}(t)\right)\right\}_{k=1,\ldots,K}.
\] 

Programs were made using R programm (see (R Development Core Team, 2005)) and are avalaible on request.
\end{itemize}
\vskip 5mm

\noindent 4.2.   MULTILAYER PERCEPTRON

\par
We used a neural network with one hidden layer having $q_2$ neurons (where $q_2$ is a parameter to be calibrated). The inputs of the neural network were:
\begin{itemize}
\item[$\bullet$] For the \emph{time series}, the disjunctive form of the value of the pixel;
\item[$\bullet$] For the \emph{spatial aspect}, the weighted frequency of each type of land cover in the neighbourhood of the pixel;
\item[$\bullet$] the environmental variables.
\end{itemize}
The output was the estimation of the probabilities (\ref{geo1_modele_nnet}).

The estimation was also made in two stages:
\begin{itemize}
\item[$\bullet$] The {\em estimation step} produces the estimated weights as described in (\ref{geo1_poids_nn}) for a given number of neurons ($q_2$) and a given neighbourhood. For this step, the neural network was trained with an early stopping procedure which allows to stop the optimization algorithm when the validation error (calculated on a part of the data set) is starting to increase (see (Bishop, 1995)).\\
This step was repeated for various values of both neighbourhood and $q_2$.
\item[$\bullet$] {\em Validation step:} once an estimation of the optimal weights was given, we chose $q_2$ and the size of neighbourhood, as for the previous model. Moreover, in order to escape the local minima during the training step, we trained the perceptrons many times for each value of neighbourhoud and of $q_2$ with various training sets; the ``best'' perceptron was then chosen according to the minimization of the validation error among both the values of the parameters (size of the neighbourhoud and $q_2$) and the optimization procedure results.
\end{itemize}

Programs were made using Matlab (Neural Networks Toolbox, see (Beale and Demuth, 1998)) and are avalaible on request.
\vskip 5mm

\noindent 5.   COMPARISON AND DISCUSSION

\par
The validation step led to select the following parameters (Table \ref{Table1}):
\begin{table}[h]
\begin{center}
\caption{Parameters selected by the validation step\label{Table1}}
\begin{tabular}{|l|c|c|}
\cline{2-3}
\multicolumn{1}{c|}{} & {\bf \textsc{Garrotxes}} & {\bf \textsc{Alta Alpujarra}}\\
\cline{2-3}
\multicolumn{3}{c}{}\\
\hline
{\bf Poly. regression} & &\\
Size of neighbourhood & 9 & 1 \\
$\epsilon$ & 10 & 0.1\\
\hhline{===}
{\bf ML perceptron} & &\\
Size of neighbourhood & 7 & 4 \\
$q_2$ & 8 & 30 \\
perceptron size & 19-8-7 & 35-30-9\\
\hline
\end{tabular}
\end{center}
\end{table}

After the two models were trained, we built the predicting map on date 2000 (Garrotxes data set) and 2001 (Alta Alpujarra data set). The performances of the two models were compared with a GIS approach.

\begin{figure}[h]
\makebox[2 cm]{}
\makebox[3.5 cm][r]{\includegraphics[width=3.5 cm]{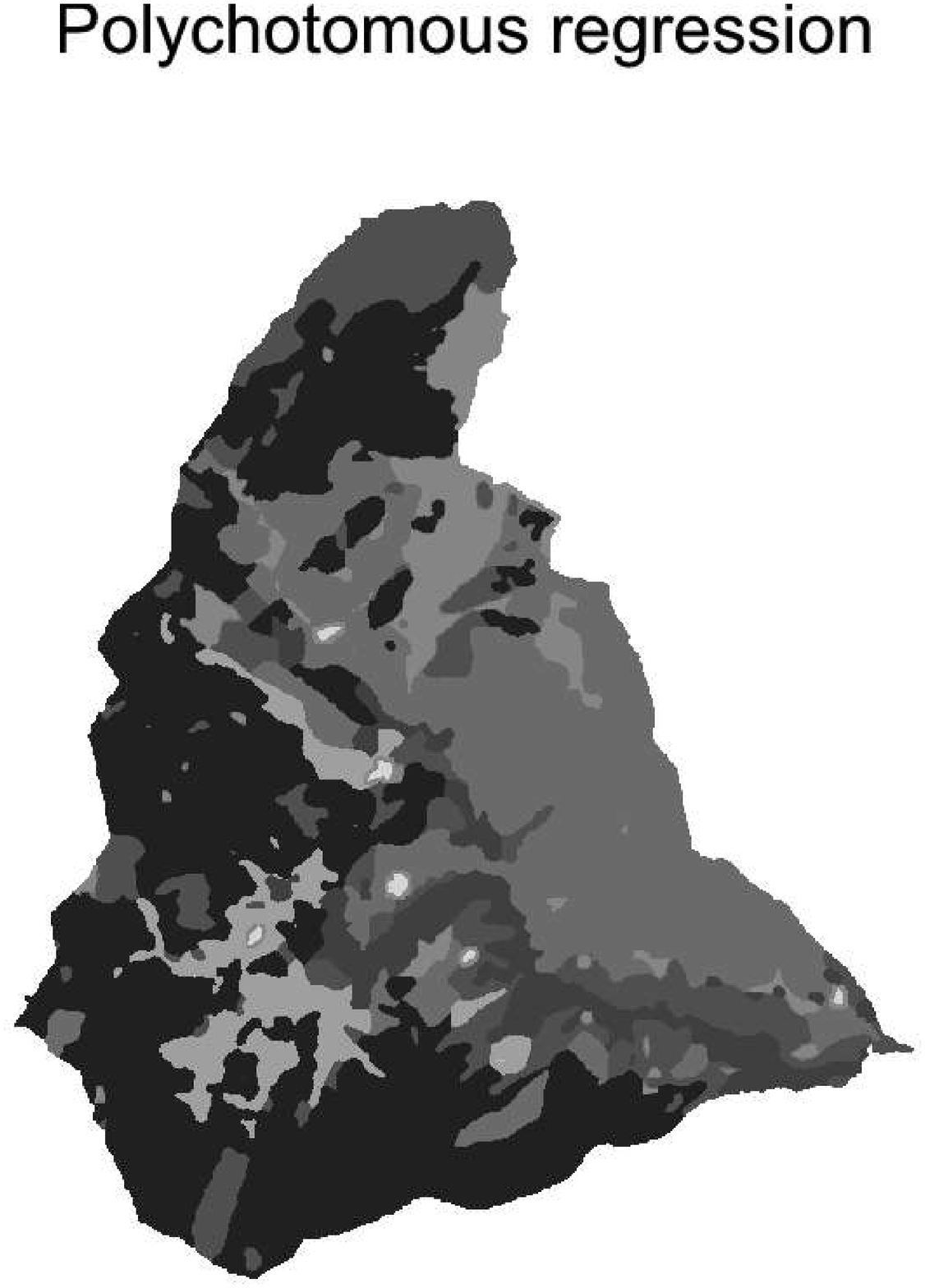}}
\makebox[1 cm]{ }
\makebox[3.5 cm][l]{\includegraphics[width=3.5 cm]{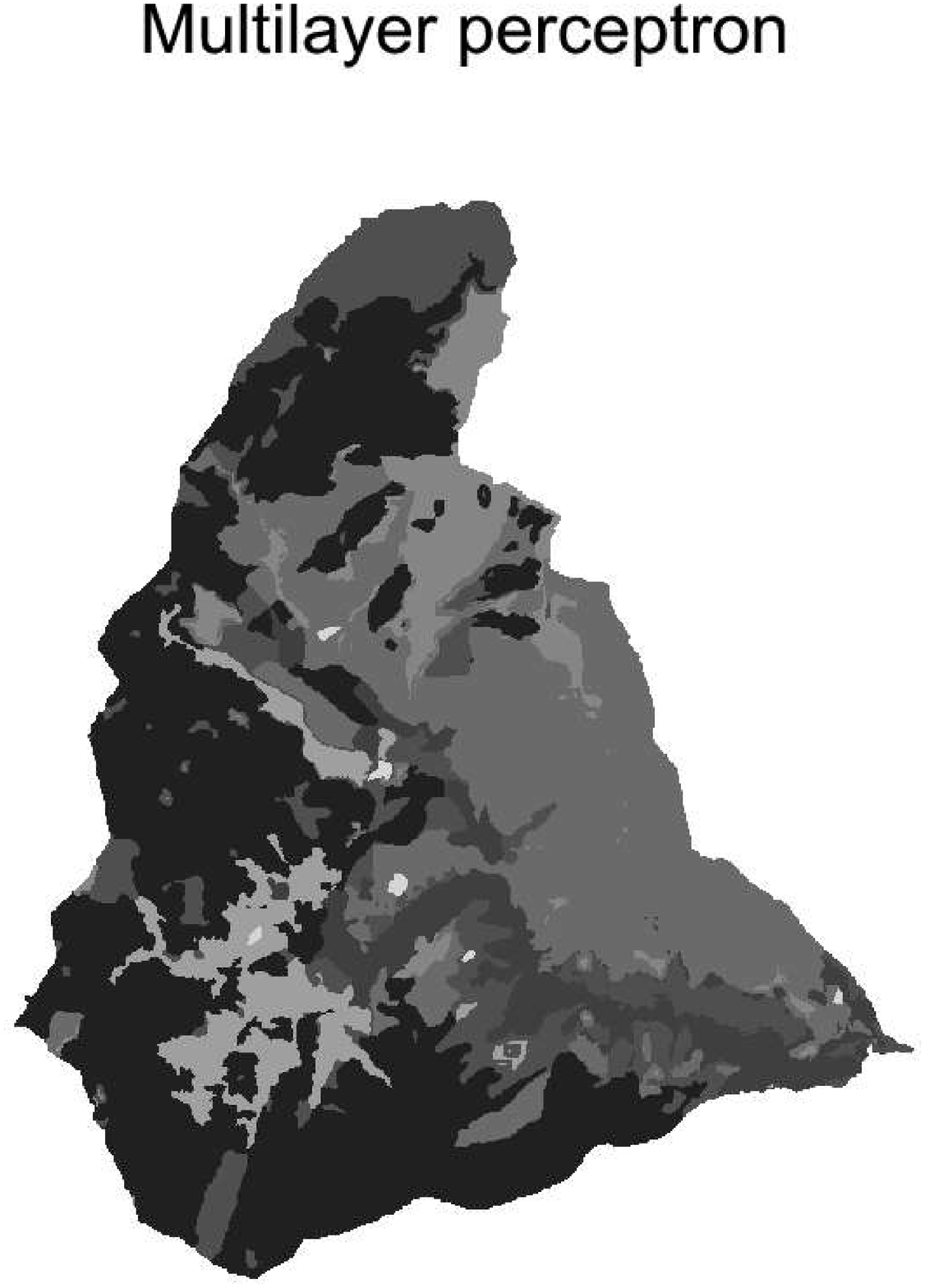}}\\
\makebox[4 cm]{}\\
\makebox[2 cm]{}
\makebox[3.5 cm][r]{\includegraphics[width=3.5 cm]{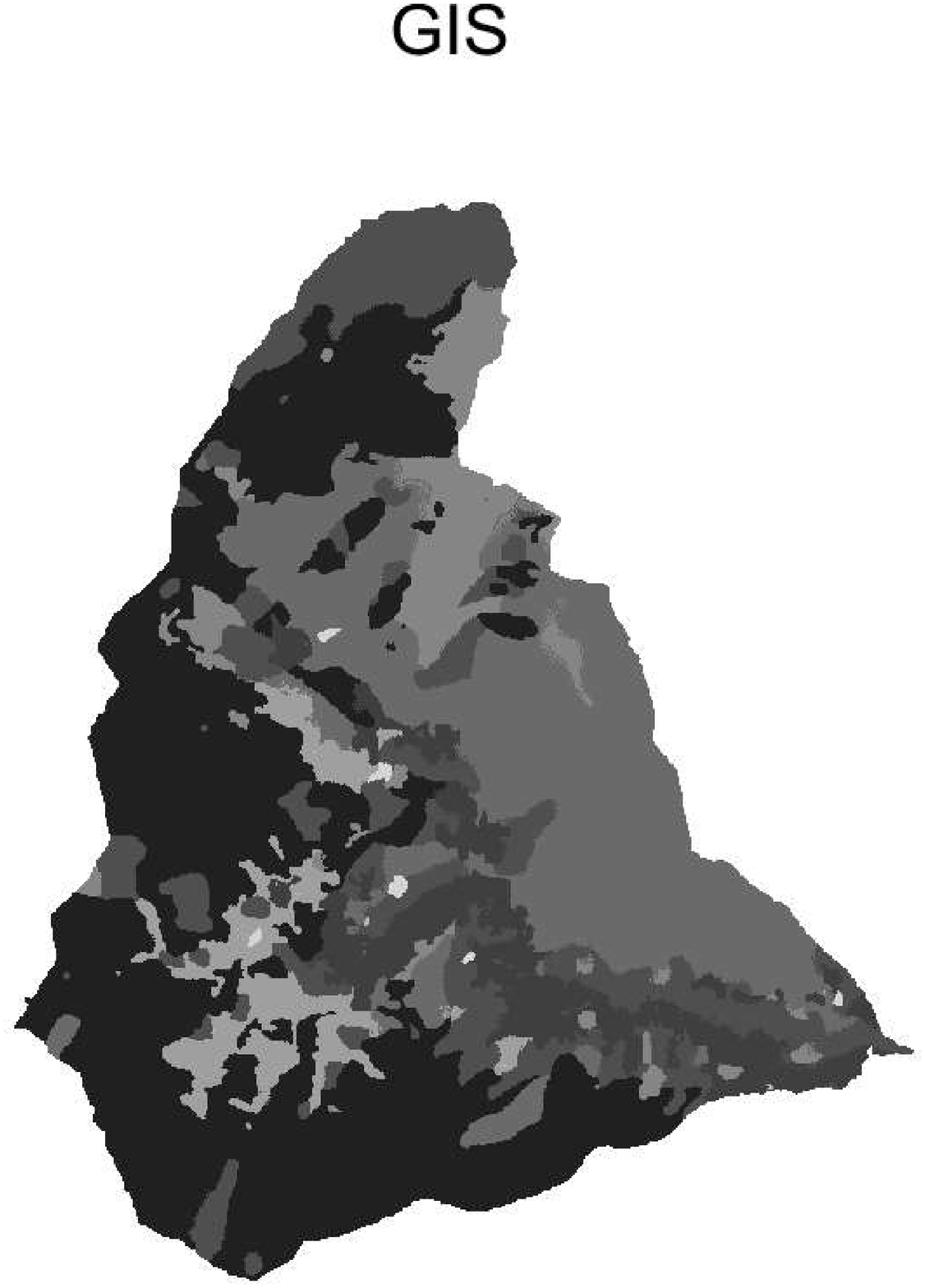}}
\makebox[1 cm]{ }
\makebox[3.5 cm][r]{\includegraphics[width=3.5 cm]{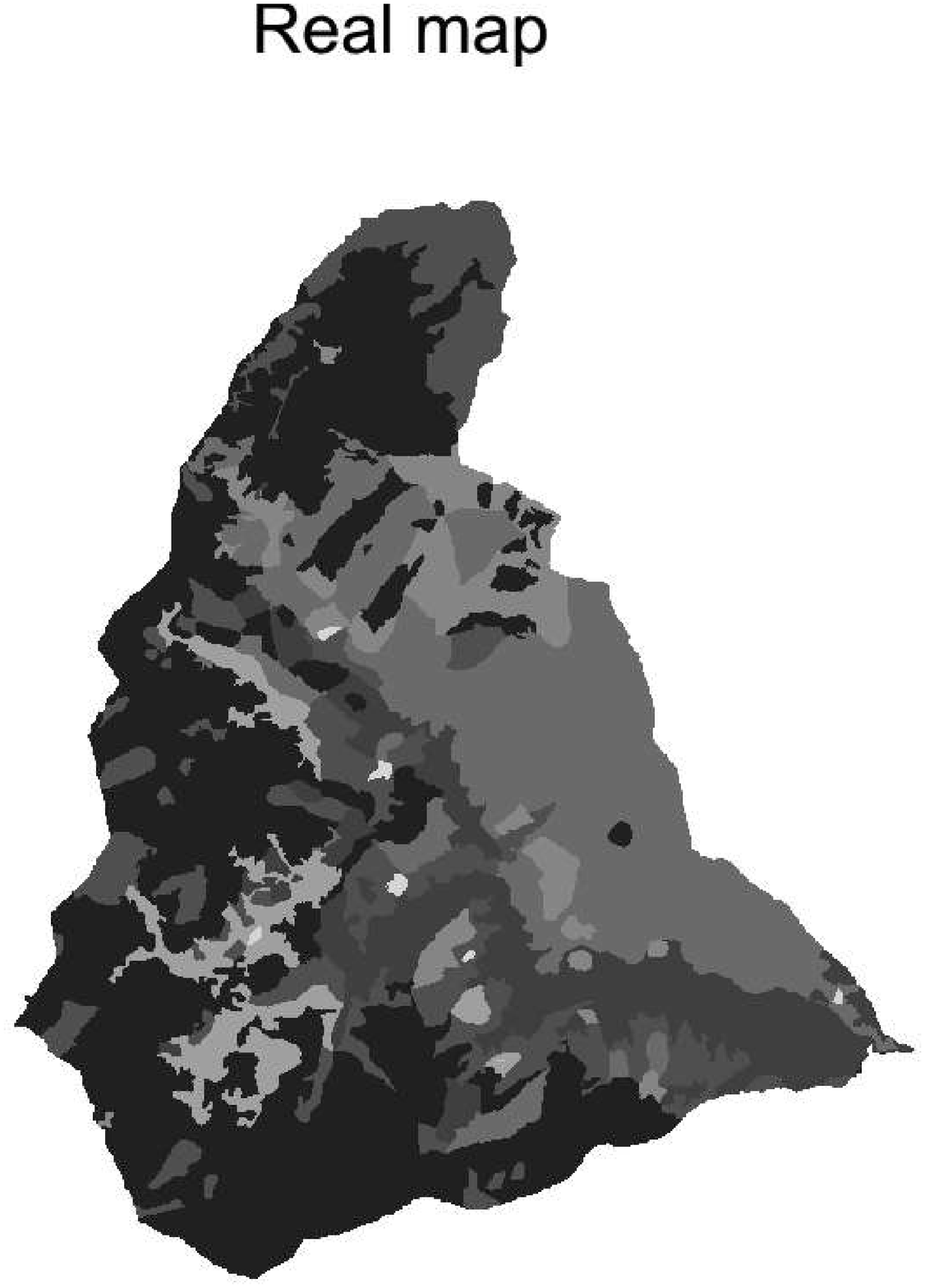}}
\caption{Predictive maps for the various approaches on date 2000 and real map (bottom right)}
\label{Figure6}
\end{figure}

For the Garrotxes data set, the results are summarized in Table \ref{Table2} and the frequencies of errors for each land cover type were calculated on the pixels which are really of this land cover type. We focus on the 6 more frequent land cover types, since the number of agriculture pixels tends to zero. In Figure \ref{Figure6}, we can see the three predictive maps given by our approaches and the GIS approach that can be confronted with the real map.
\begin{table}[h]
\begin{center}
\caption{Missclassification rates for the Garrotxes}
\label{Table2}
\begin{tabular}{|c|c|c|c|c|}
\hline
Land cover & Frequency & Poly. Regression & ML perceptron & GIS \\
types & in the area & error rate	& error rate & error rate \\
\hline
Coniferous forests & 40.9 \% & 11.9 \% & 10.6 \% & 11.4 \%\\
Deciduous forests & 11.7 \% & 51.7 \% & 45.8 \% & 55.3 \%\\
Scrubs & 15.1 \% & 57.1 \% & 54.5 \% & 51.9 \%\\
Broom lands & 21.6 \% & 14.4 \% & 16.2 \% & 17.1 \%\\
Grass pastures & 5.7 \% & 59.2 \% & 59.4 \% & 54.4 \%\\
Grasslands & 4.8 \% & 25.6 \% & 19.3 \% & 30.4 \%\\
\hline
\textbf{Overall} & & \textbf{27.2 \%} & \textbf{25.7 \%} & \textbf{27.2 \%}\\
\hline
\end{tabular}
\end{center}
\end{table}

\begin{figure}[h]
\makebox[2 cm]{}
\makebox[3.5 cm][r]{\includegraphics[width=3.5 cm]{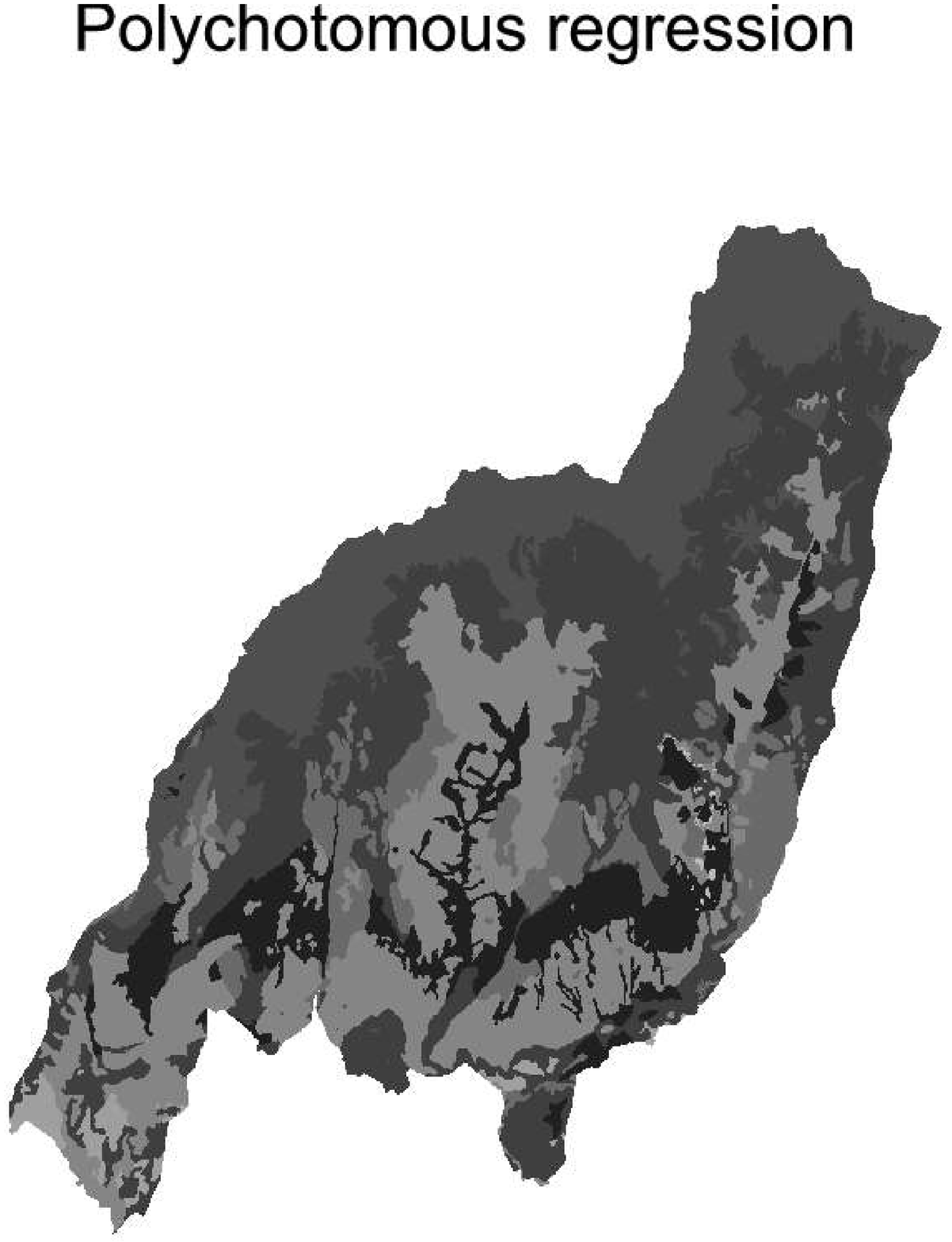}}
\makebox[1 cm]{ }
\makebox[3.5 cm][l]{\includegraphics[width=3.5 cm]{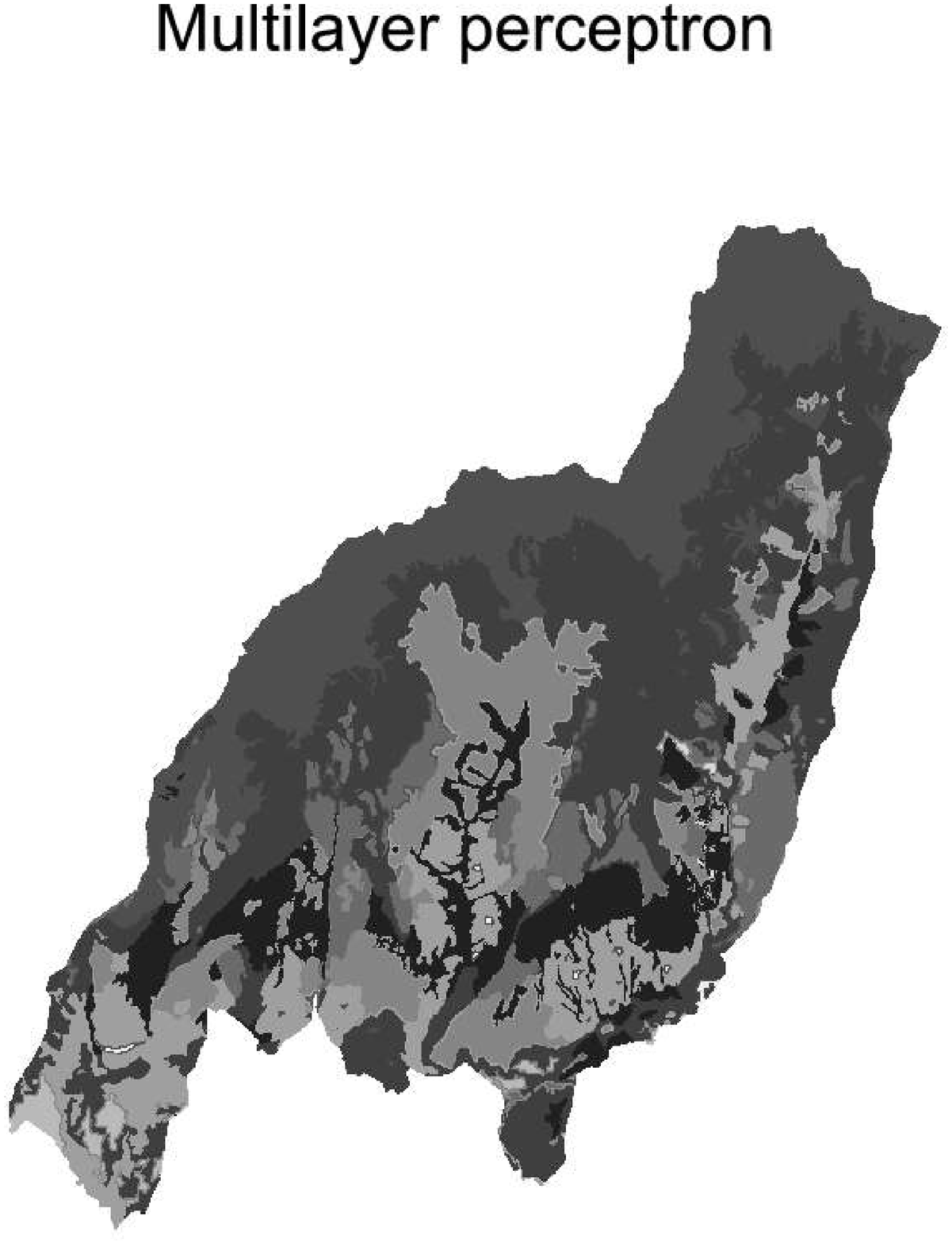}}\\
\makebox[4 cm]{}\\
\makebox[2 cm]{}
\makebox[3.5 cm][r]{\includegraphics[width=3.5 cm]{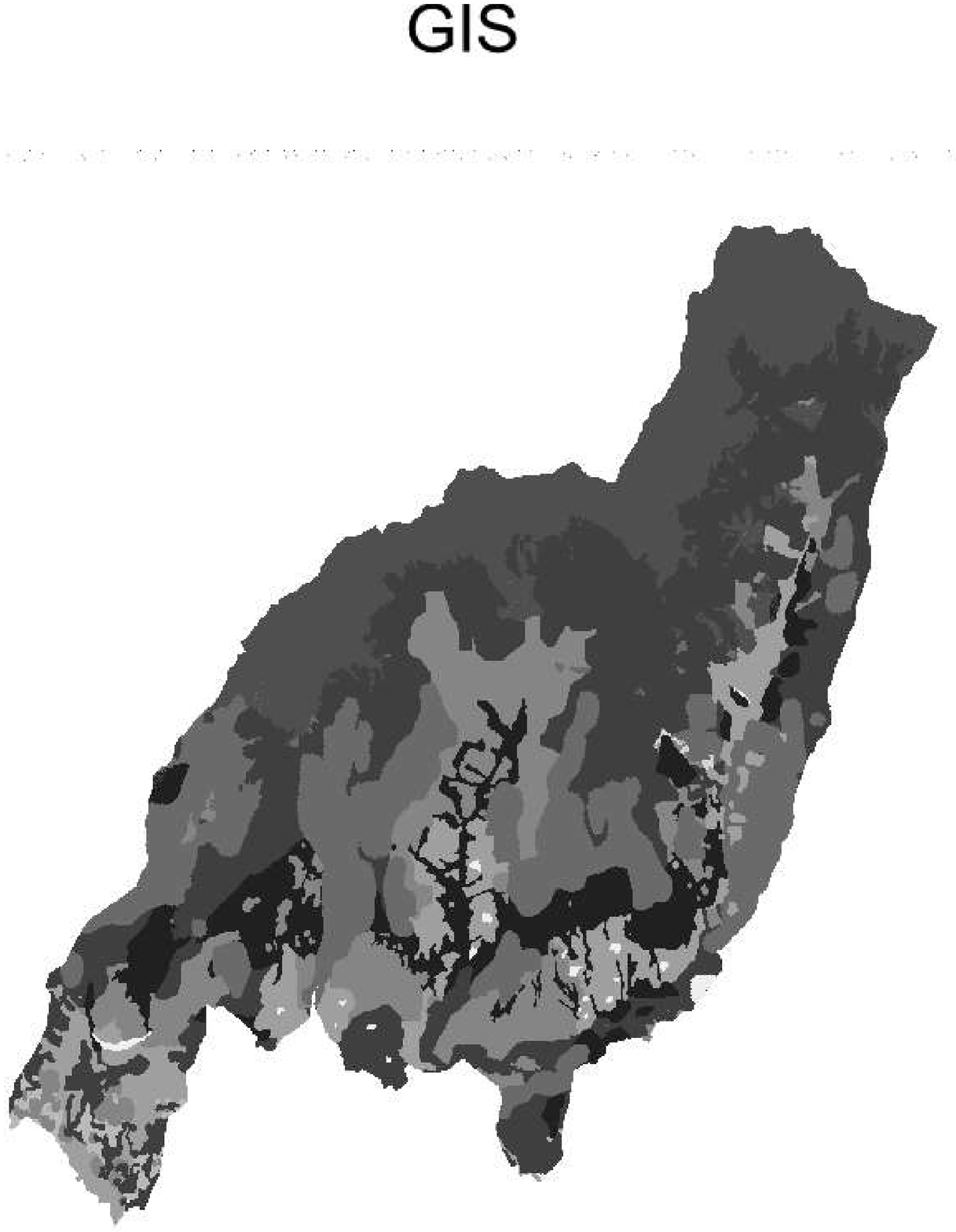}}
\makebox[1 cm]{ }
\makebox[3.5 cm][r]{\includegraphics[width=3.5 cm]{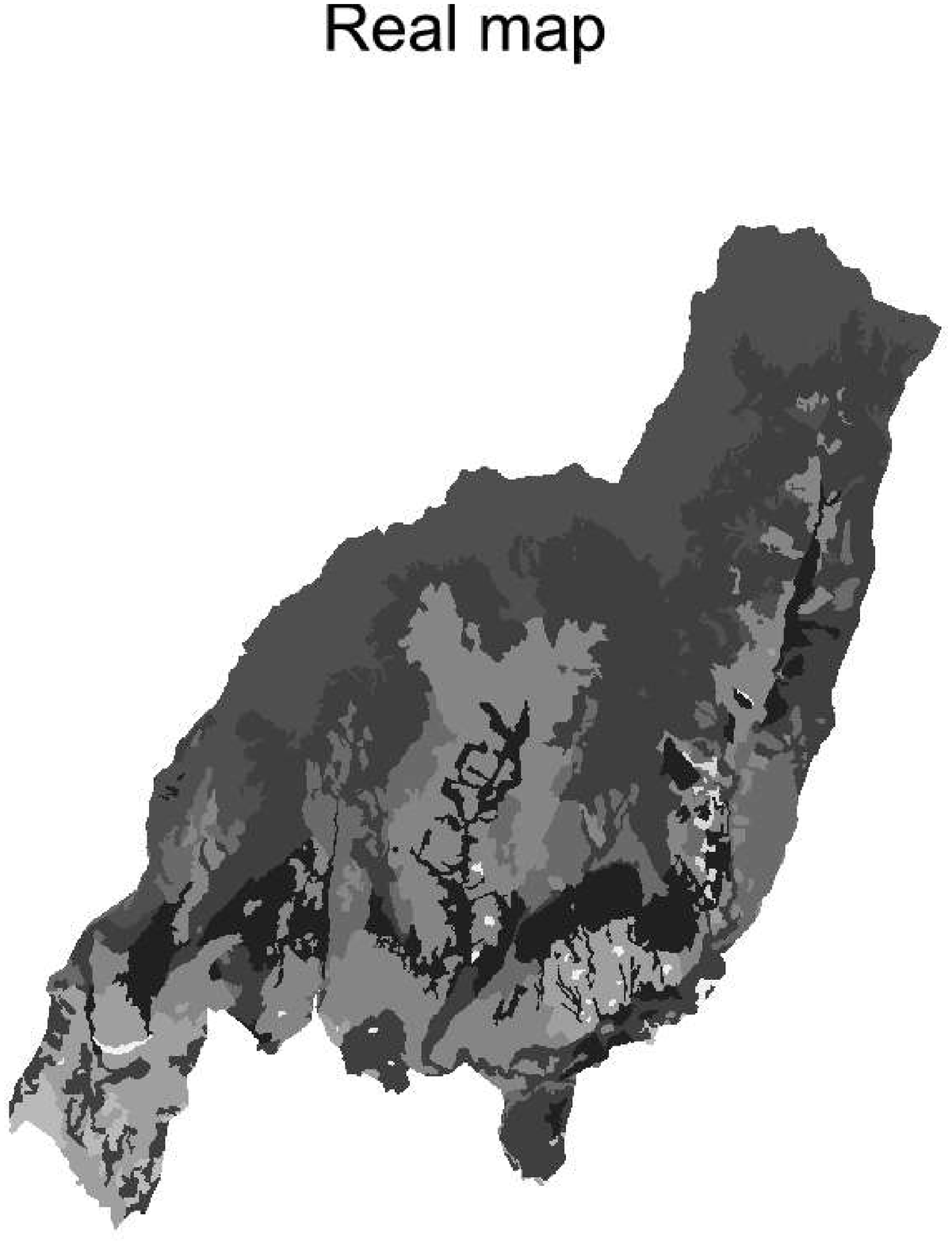}}
\caption{Predictive maps for the various approaches on date 2001 and real map (bottom right)}
\label{Figure7}
\end{figure}

For the Alta Alpujarra data set, the results are summarized in Table \ref{Table3} (land cover type under 5 \% of the area have been omitted). Predicted maps and real maps are compared in Figure \ref{Figure7}.
\begin{table}[h]
\begin{center}
\caption{Missclassification rates for the Alta Alpujarra}
\label{Table3}
\begin{tabular}{|c|c|c|c|c|}
\hline
Land cover & Frequency & Poly. Regression & ML perceptron & GIS \\
types & in the area & error rate	& error rate & error rate \\
\hline
Deciduous forests & 10.9 \% & 3.5 \% & 2.6 \% & 14.3 \%\\
Scrubs & 33.0 \% & 3.1 \% & 1.4 \% & 15.2 \%\\
Pasture & 20.8 \% & 0.6 \% & 0 \% & 12.5 \%\\
Coniferous refor. & 9.23 \% & 3.5 \% & 16.3 \% & 1.9\%\\
Fallows & 18.8 \% & 32.5 \% & 41.4 \% & 46.8\%\\
Irrigated cropland & 5.8 \% & 8.9 \% & 6.8 \% & 38.9\%\\
\hline
\textbf{Overall} & & \textbf{9.0 \%} & \textbf{11.28 \%} & \textbf{21.1 \%}\\
\hline
\end{tabular}
\end{center}
\end{table}

First of all, the predictive maps provided by the two statistical methods are coherent, smooth and close to reality. This can also be shown through the good error rates (about 25 \% - 27 \% for the Garrotxes data set and 9 \% - 12 \% for the Alta Alpujarra) which are clearly a good performance considering the poverty of the data (we only had got 3 or 4 dates to train the models).\\
Furthermore, the striking fact is that the ``automatic'' statistical approaches did as well (Garrotxes data set) or even much better (Alta Alpujarra) than the guided GIS approach. This is an interesting point in order to help improving the classical geographical approach to predicting land cover, and better understand the environmental changes in time and space. Moreover, the ``automatic'' statistical methods were much faster than the GIS as they do not use any expert knowledge which takes a long time to be modelized and needs to be remade for each area. On the contrary, the polychotomous regression modelling and the multilayer perceptron approaches did not lead on these data sets to significant differences. The first method was much faster to train and it was then quite attractive to use it. However, we think that, on a general point of view, the greater flexibility of multilayer perceptron could be usefulness to predict land cover for other data sets where a parametric model could fail.

The main advantage of the automatic statistical approaches is in the fact that they simultaneously take into account the spatio-temporal aspect of the problem and also the environmental variables. GIS works in two steps: it first predicts the number of pixels for each land cover type by a simple temporal model and then takes into account the spatial aspect and the environmental variables to allocate these pixels spatially. This could partially explain that GIS had worse performances for the Alta Alpujarra data set, as the coniferous reforestation used to be important in the 60's and has then be given up. This led the GIS to predict, in the 2001 map, much more coniferous reforestation pixels than in the real map: 18.8 \% of the pixels were predicted in the coniferous reforestation type against 7.9 \% for the multilayer perceptron, 9.6 \% for the polychotomous regression modelling and 9.2 \% for the real map. Then GIS approach had a much lower error rate on the coniferous reforestation land cover type but a bigger one for the other ones.

Finally, looking further in the missclassification rates for the various land cover types, we can see that the most dynamic land cover type were harder to train: this is the case, for instance, for the scrubs in the Garrotxes area where they tended to grow fast and became deciduous forests; this is also the case, in the Alta Alpujarra for the fallows and irrigated croplands because agricultural lands were tending to be left. These dynamics could be better predicted by adding pertinent informations for these kinds of land cover types (density of the scrubs, for example, can help knowing if they can, or not, become forests). 
\vskip 5mm

\noindent 6.   CONCLUSION

\par
Finally, this work shows the great potential of the two statistical models in predictive prospection on geographical data. These models had as good performances as GIS approach and we can hope that a combination of the two points of view (statistics and GIS) can improve the land cover predictions: the empirical first step of the GIS could be improved by being replaced by one of these statistical approaches. This issue, that is of big interest for geographers, is still under study as the GIS approach was performed through pre-made programs and has then to be totally re-though to that aim.

Another aspect that has to be worked on is the form of the data: for example, we underlined that an information on the density of the scrubs is needed to better understand their evolution. This could help geographers to better understand what is of interest for predicting the land cover evolution for their future studies.
\vskip 5mm

\noindent 7.   ACKNOWLEDGEMENTS

\par
The authors are grateful to the Ministerio de Cencia y Tecnologia who supports this research (Plan nacional de investigacion cientifica, Desarollo e innovacion tecnologica, BIA 2003\_01499).

The authors are also grateful to the anonymous referees for their detailed and constructive comments and suggestions which have substantially improved the manuscript.

\vskip 6mm
\baselineskip=12pt

\noindent BIBLIOGRAPHY
\vskip 3mm

\noindent (1) Beale, M., Demuth, H. (1998). {\it Neural network toolbox user's guide}. Version 3. The Matworks Inc.

\noindent (2) Bishop, C. (1995). {\it Neural Networks for Pattern Recognition}. New York: Oxford University Press.

\noindent (3) Cardot, H., Faivre, R., Goulard, M. (1993). Functional approaches for predicting land use with the temporal evolution of coarse resolution remote sensing data. {\it Journal of Applied Statistics}, 30: 1185--1199.

\noindent (4) Hornik, K. (1991). Approximation capabilities of multilayer feedforward networks. {\it Neural Networks}, 4(2): 251--257.

\noindent (5) Hosmer, D., Lemeshow, S. (1989). {\it Applied logistic regression}. New York: Wiley.

\noindent (6) Kooperberg C., Bose, S., Stone, J. (1997). Polychotomous regression. {\it Journal of the American Statistical Association}, 92: 117--127.

\noindent (7) Lai, T. and Wong, S. (2001). Stochastic neural networks with applications to nonlinear time series. {\it Journal of the American Statistical Association}, 96(455): 968--981.

\noindent (8) R Development Core Team (2005). {\it R: a language and environment for statistical computing}. Vienna, Austria: R Foundation for Statistical Computing.

\noindent (9) Paegelow, M., Camacho Olmedo M.T. (2005). Possibilities and limits of prospective GIS land cover modeling - a compared case study: Garrotxes (France) and Alta Alpujarra Granadina (Spain). {\it International Journal of Geographical Information Science}, 19(6): 697--722. 

\noindent (10) Paegelow, M., Villa, N., Cornez, L., Ferraty, F., Ferré, L. and Sarda, P. (2004). Modélisations prospectives de l'occupation du sol. Le cas d'une montagne méditerranéenne. {\it Cyberg\'eo}, 295.

\noindent (11) Parlitz, U., Merkwirth, C. (2000). Nonlinear prediction of spatio-temporal time series. In {\it ESANN'2000 proceedings}. Bruges, Belgium: 317--322.

\noindent (12) White, H. (1989). Learning in artificial neural network: a statistical perspective. {\it Neural Computation}, 1: 425--464.

\end{document}